\documentclass[12pt,tightenlines]{revtex4-2}
\usepackage{amsfonts,amsmath,amssymb}
\usepackage{txfonts}
\usepackage{graphicx,color}
\usepackage{wrapfig,lipsum}
\usepackage{floatrow,float,booktabs}
\usepackage{epstopdf}
\usepackage{hyperref}
\usepackage[normalem]{ulem}
\usepackage{enumitem}
\usepackage{braket}
\usepackage{mathrsfs,mathtools}
\usepackage{cancel}
\usepackage{lineno}
\newcommand{\so}{\operatorname{SO}}
\begin{document}
	\title{Neural Receptive Fields, Stimulus Space Embedding and Effective Geometry of Scale-Free Networks}
	\author{Vasilii Tiselko$^{1,2}$, Alexander Gorsky$^{1,2,3}$, Yuri Dabaghian$^{4}$}
	\affiliation{
		$^1$\small Laboratory of Complex Networks, Center for Neurophysics and Neuromorphic Technologies, Moscow, Russia,\\
		e-mail: vasily.tiselko@gmail.com \\
		$^2$\small Moscow Center for Advanced Studies, Moscow, Russia,\\
		$^3$\small Institute for Information Transmission Problems RAS, Moscow, Russia,\\
		$^4$\small Department of Neurology, The University of Texas Health Science Center at Houston, Houston, USA\\
		e-mail: yuri.a.dabaghian@uth.tmc.edu
	}
\begin{abstract}
	Understanding how receptive fields emerge and organize within brain networks and how neural dynamics couple with stimuli space is fundamental to neuroscience. Models often rely on fine-tuning connectivity to match empirical data, which may limit biological plausibility. Here we propose a physiologically grounded alternative where receptive fields and population-level attractor dynamics arise naturally from the effective hyperbolic geometry of scale-free networks. By associating stimulus space with the boundary of a hyperbolic embedding, we simulate neural dynamics using rate-based and spiking models, revealing localized activity patterns that reflect stimulus space structure without synaptic fine-tuning. The resulting receptive fields follow experimentally observed statistics and properties, and their sizes depends on neuron's connectivity degree. The model generalizes across stimuli dimensionalities and various modalities, such as orientation and place selectivity. Experimental analyses of hippocampal place fields recorded on a linear track support these findings. This framework offers a novel organizing principle linking network structure, stimulus space encoding, and neural dynamics, providing insights into receptive field formation across
	diverse brain areas.
\end{abstract}
\maketitle
\newpage

\section{Introduction}
\label{sec:int}

One of the foundational questions in neuroscience concerns how neural activity becomes coupled to specific stimulus processing, how internal network dynamics relate to external variations, and how this coupling is grounded in structural organization.
Across different brain regions, from primary sensory cortices to deep networks such as the hippocampus, neural responses exhibit selectivity to specific stimuli. The phenomenon is so pervasive that it is often taken as a reflection of some fundamental principle of information processing within brain networks. Yet mechanistic understanding remains largely speculative even for stimuli that are readily interpretable, as in the case of orientation encoding or spatial navigation.

Traditional models often rely on extensive fine-tuning of connectivity so that neural dynamics matches the empirically observed structure of the stimulus or responses. However, such approaches limit biological plausibility and fail to explain complex stimuli and diverse dynamical phenomena. Here we propose a different approach, introducing a distinct class of mechanisms that avoids $ad$ $hoc$ constraints on local structural plasticity and is instead grounded in geometric properties of network organization at a more global level---in the statistics governing how connectivity is distributed across neurons.

\begin{figure}[]
	\centering
	\includegraphics[width=0.84\textwidth]{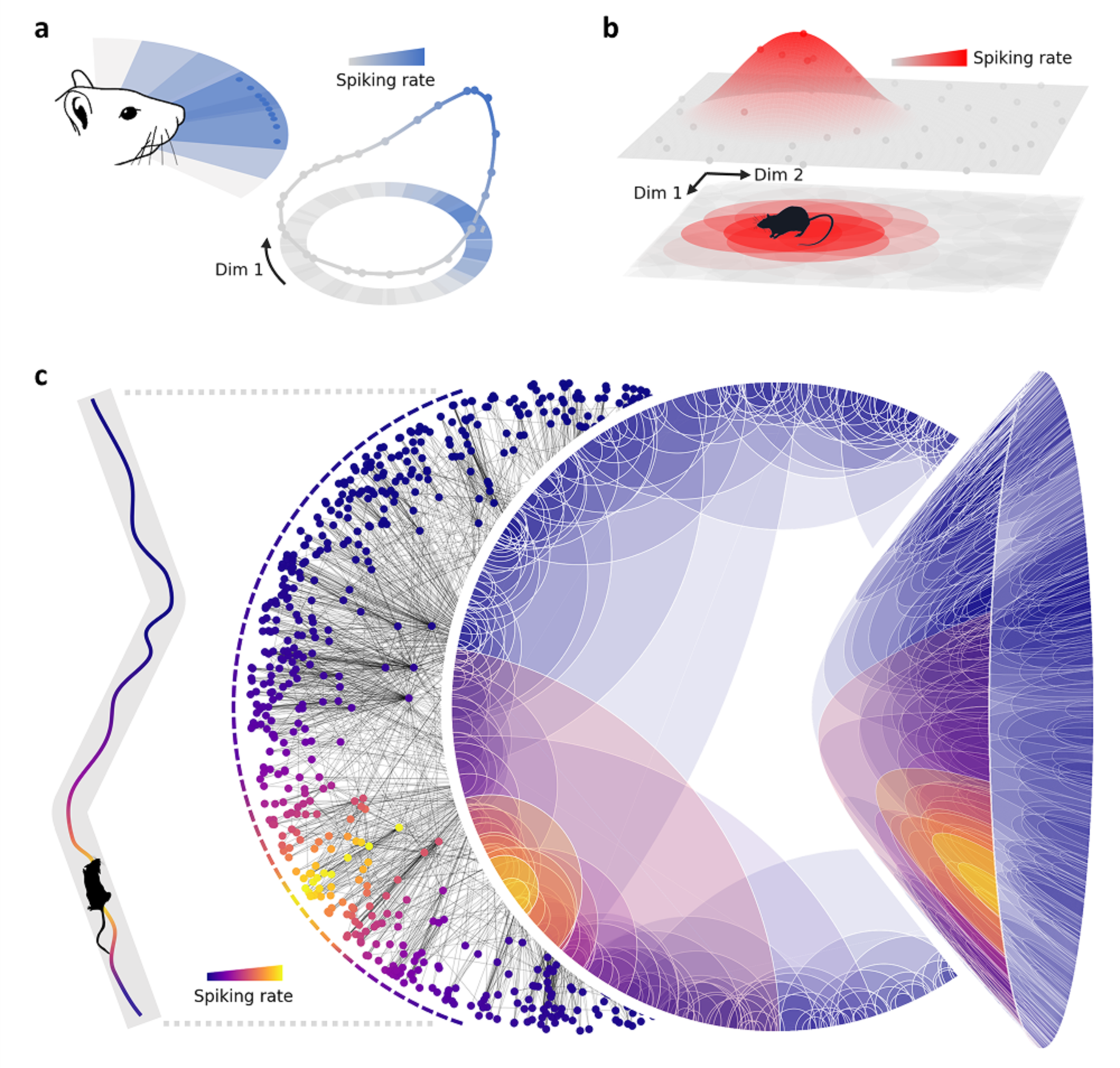}
	\caption{\footnotesize{
			\textbf{Neural responses are governed by the receptive fields.}
			\textbf{(a)} A map of angular domains---receptive fields of head-direction cells---covering configurational space of orientations, \textit{i.e.}, a topological cycle. (Blue dots schematically represent the firing of individual head-direction neuron, similar to the place cell responses illustrated in panel b.)
			\textbf{(b)} Hippocampal place cells spike when the animal appears in particular regions in the navigated environment (red circular domain). Place fields tile the navigated environment and spiking is localized on neurons whose receptive fields intersect with the animal's current position in space, thus reflecting the current stimulus state.
			\textbf{(c)} The proposed approach enables a principled coupling between stimulus spaces and network dynamics without compromising structural flexibility. The stimulus space can be associated with an effective boundary space of the network’s hyperbolic embedding due to the geometric properties of scale-free networks. Receptive-field organization inherits this hyperbolic geometry, and the resulting family of localized activity states mirrors the structure of the stimulus space.
	}}
	\label{f1:receptive_fields}
\end{figure}

Neural spiking responses across various stimuli are reflected by receptive fields --- specific domains within a stimulus space that elicit activity from individual cells (Fig.~\ref{f1:receptive_fields}). Receptive fields, which vary in size, shape, and dimensionality, are observed not only in sensory systems (visual~\cite{hubel1959receptive,
	girman1999receptive, bednar2003self}, auditory~\cite{aertsen1981spectro}, olfactory~\cite{Imi}, \textit{etc}.) but also in higher-order associative networks such as the hippocampus~\cite{mehta2000experience,rolls1997spatial}, postsubiculum~\cite{taube1990head,taube1998head}, and entorhinal cortex (Fig.~\ref{f1:receptive_fields}a). Collectively, the arrangement of receptive fields reflects the structure of the stimulus space and links it to neural dynamics.

In sensory cortices, proximity in stimulus space is often mirrored anatomically: neighboring neurons tend to exhibit adjacent or overlapping receptive fields, as observed in orientation-selectivity maps of the visual cortex or tonotopic maps of the auditory cortex, and so forth \cite{Ptl,Srn}. As a result, neurons with neighboring receptive fields form a localized ``bump" of activity that shifts coherently with stimulus changes.

The physiological basis for such behavior is thought to arise from the adaptation of neuronal connectivity to the geometry of the stimulus space manifold~\cite{bednar2003self, sirosh1996self}. In particular, the phenomenon of localization is attributed to attractor dynamics---nonlinear interactions between excitatory and inhibitory neuronal populations that stabilize activity over timescales much longer than individual spikes~\cite{tiselkoapparent, hulse2020mechanisms,bassett2018self, laurens2018brain}. This reasoning forms the foundation of most Continuous Attractor Neural Network (CANN) models, which achieve coherent attractor states through a combination of local excitation and long-range or homogeneous inhibition \cite{Trppbg,Smsch,RllAtt,Wbr,WuZhang}.

Importantly, the notion of ``locality" in these models is based on either the physical proximity between neurons or the distances observed between receptive fields~\cite{romani2010continuous}. Although this approach can account for certain population-level features of neuronal dynamics, including the localization of spiking, it remains phenomenological, \textit{i.e.}, based on external assessments and tuning of local connectivity to capture the structure of the stimulus space, and may therefore misrepresent the intrinsic neurophysiological computations \cite{Trppbg,RllEp,RllAtt,ClgAtt,Rll4,Bsstt,StrngHD,StrngSV,SkgDir,Smsch,Kali,Wbr,Wlls,Tsdks,WuZhang}. In particular, this methodology struggles in the case of the hippocampus---a deep brain network whose principal neurons, place cells, fire only within specific regions of the environment, known as their respective place fields (Fig.~\ref{f1:receptive_fields}b).
%%%%%%%%%%%%%%%%%%%%%%%%%%%%%%%%%%%%%%%%%%%%%%%%%%%%%%%%%%%%%%%%%%%%%

%%%%%%%%%%%%%%%%%%%%%%%%%%%%%%%%%%%%%%%%%%%%%%%%%%%%%%%%%%%%%%%%%%%%%
First, this network integrates inputs from a wide array of sensory, vestibular, motor, and associative cortices, all of which are heavily processed and interwoven within a highly intricate anatomical organization and cannot be naively metrized~\cite{hartley2000modeling, araujo2001view, blair1996visual,knierim1995place, knierim1998interactions,gerlei2020grid}. 

Second, place field's localization in the environment is largely independent from the place cells' physical positions \cite{Moser2014}. Third, hippocampus exhibits a number of complex transient dynamics, \textit{e.g.}, new maps are generated upon exposure to each novel environment. Yet, in each map, the ongoing firing activity is localized on neurons whose place fields overlap at the animal's current location, while other cells remain suppressed, which also points at the presence of a localized activity bump.

Indeed, such a bump can be constructed by rearranging place cells into an auxiliary topographic configuration (Fig.~\ref{f1:receptive_fields}c). This suggests that hippocampal activity may also be governed by CANN dynamics, provided that one can identify a biologically plausible network connectivity with sufficient structural flexibility, define an appropriate metric on the incoming stimuli, and establish a mechanism for extracting their proximities in a manner consistent with externally observed parameters.

To address these problems, we propose an approach that circumvents the need for \textit{ad hoc} adjustments of local connectivity, while enabling a principled coupling between stimulus spaces and network dynamics without compromising structural flexibility. The stimulus space can be associated with an effective embedding boundary space due to the geometric properties of scale-free networks---an architecture ubiquitous across biological systems, from whole brain connectomes~\cite{tadic2019functional, zheng2019geometric, whi2022hyperbolic, allard2018navigable} to neural circuits~\cite{li2010scale, lynn2024heavy}. A defining feature of these networks is their hierarchical connectivity, characterized by a power-law degree distribution that gives rise to hyperbolic geometric effects \cite{krioukov2010hyperbolic,Bgn}. We use these properties to demonstrate that spiking dynamics in such networks naturally generate stable activity localization that responds adaptively to external inputs and produces complex yet faithful tilings of the stimulus space. These tilings are characterized by exponentially distributed receptive field sizes, consistent with empirical observations \cite{zhang2023hippocampal}. 

Taken together, this model offers a framework for how scale-free network structure may support the coupling with the stimulus spaces and its partition with receptive fields, as demonstrated in our simulations and analyses of hippocampal place fields.

\section{Approach: hyperbolicity and stimulus space embedding}
\label{sec:app}

\begin{figure}
	%\begin{wrapfigure}{c}{0.75\textwidth}
	\centering
	\includegraphics[width=0.95\textwidth]{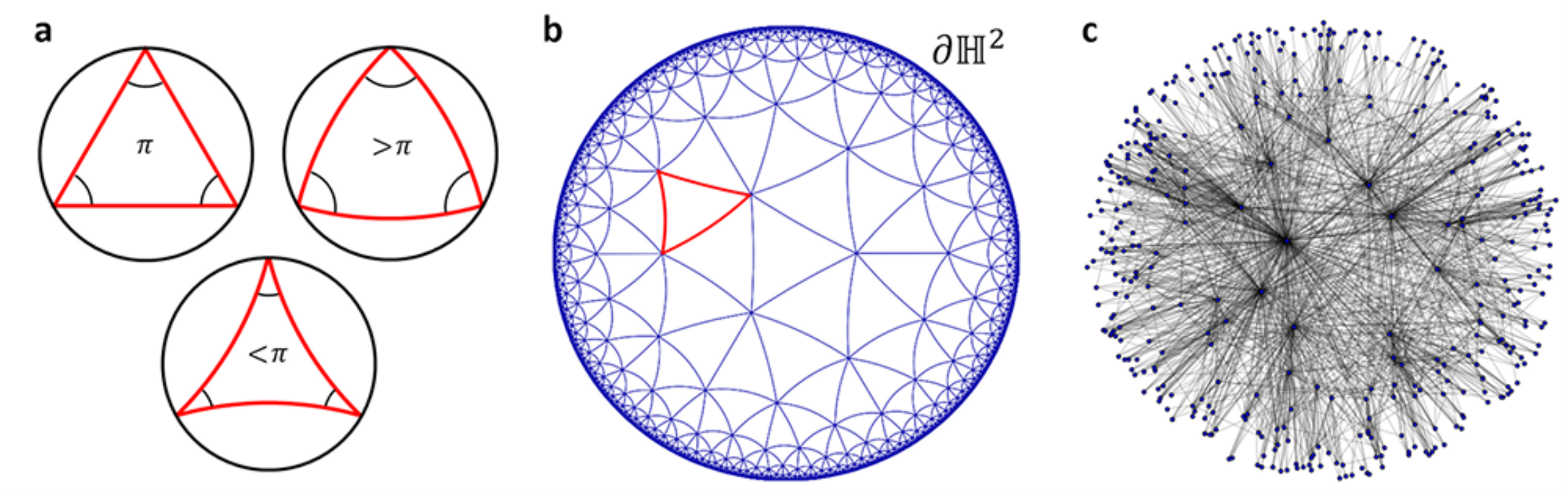}
	\caption{\footnotesize{\textbf{Hyperbolic spaces and scale-free networks.}
			\textbf{(a)} Three isotropic $2D$ geometries: flat Euclidean space (left), spherical space with positive curvature (right), and hyperbolic space with negative curvature (bottom).
			\textbf{(b)} Poincar\'e model: the entire $2D$ Lobachevsky plane, $\mathbb{H}^2$, is mapped conformally onto the compact Euclidean disk. The boundary of the disk, corresponding to $S^1$, represents points at infinity, enclosing $\mathbb{H}^2$. Geodesic segments (blue lines) form angle-deficient triangles that tile the entire $\mathbb{H}^2$ space.
			\textbf{(c)} Scale-free networks exhibit an effective hyperbolic structure that allows embedding them within the Poincar\'e disk model. Each node $v$ is assigned a radial coordinate (embedding radius) reflecting its degree---nodes with higher connectivity (hubs) are positioned closer to the center, while weakly connected nodes are pushed toward the boundary. Smaller angular distance between nodes reflects similarity in local connectivity structure and a higher probability of being connected.
	}}
	\label{f2:hyperbolic_embedding}
	%\end{wrapfigure}
\end{figure}

Hyperbolic geometry underlies a wide range of phenomena across scientific disciplines. One of its key properties is a negative curvature, which distinguishes hyperbolic spaces from flat Euclidean and positively curved spherical geometries (Fig.~\ref{f2:hyperbolic_embedding}a).

A canonical representation of the simplest, two-dimensional ($2D$) hyperbolic space $\mathbb{H}^2$ is the Poincar\'e model---a conformal embedding of the entire $\mathbb{H}^2$ into the Euclidean disk (Fig.~\ref{f2:hyperbolic_embedding}b). This construction preserves the angular structure but distorts the distances. As depicted in Fig.~\ref{f2:hyperbolic_embedding}, the effective volume of the hyperbolic plane $\mathbb{H}^2$ concentrates towards the boundary, which is reflected by the increasing density of points whose originals are uniformly distributed throughout $\mathbb{H}^2$. The $\mathbb{H}^2$-geodesics form arcs orthogonal to the disk's boundary or straight diameters. The compact, one-dimensional ($1D$) boundary of the disk corresponds to ``points at infinity"---an ideal boundary of $\mathbb{H}^2$ that captures the asymptotic directions of the geodesics. This boundary plays a fundamental role in many theoretical frameworks~\cite{krioukov2010hyperbolic, serrano2007self}, including our own, as discussed below.

Recently it was shown that scale-free networks, characterized by heterogeneous power-law distributions of node degrees, naturally exhibit effective hyperbolic geometric properties ~\cite{krioukov2010hyperbolic, serrano2007self}. This effect can be manifested explicitly by embedding such a network into a hyperbolic space, where the network’s connectivity is reflected by geometric proximity (Fig.~\ref{f2:hyperbolic_embedding}c). 
%%%%%%%%%%%%%%%%%%%%%%%%%%%%%%%%%%%%%%%%%%%%%%%%%%%%%%%%%%%%%%%%%%%%%

%%%%%%%%%%%%%%%%%%%%%%%%%%%%%%%%%%%%%%%%%%%%%%%%%%%%%%%%%%%%%%%%%%%%%
In this embedding, each node acquires hyperbolic coordinates: a radius $r$ and a set of angles, $\theta_i$, whose number is defined by dimensionality of the hyperbolic model---the approach generalizes seamlessly to any dimensions $\mathbb{H}_n$). The radial coordinate $r$ is associated with the node's degree and captures the internal hierarchy on the network structure via
\begin{equation}
	k \propto e^{-r}.
	\label{eq:degree_radial}
\end{equation}
As a result, nodes with higher degrees appear near the center of the hyperbolic disk, forming a strongly connected core, while nodes with weak connectivity pulled toward the embedding boundary. The angular coordinates, $\theta_i$, represent the local connectivity between nodes through the generalized angular distance ~\cite{krioukov2010hyperbolic, serrano2007self}. In our study,  scale-free organization provides the basis for establishing excitatory neuronal connections. % (see Methods, \ref{sec:met}). 

A central idea of our approach is to associate the boundary of hyperbolic embedding of a scale-free network with the stimulus space, following \cite{krioukov2010hyperbolic,serrano2007self}. Based on the correspondence between the hyperbolic model and its boundary-related spherical representation, we assume that neurons located at angular positions $\theta_i$ in the embedding are driven by stimuli originating from the corresponding domains of the stimulus space. Structural similarity appears in many functionally distinct stimulus spaces, such as those driving orientation-selective neurons in the visual cortex, head-direction cells in the postsubiculum, or hippocampal place cells along quasilinear tracks. All these cases have topologically equivalent stimulus spaces that can be described by a $1D$ angular variable \cite{Krg,Kng,Rbk}.

With vertical orientation included, head-direction cells’ stimuli cover a $2D$ sphere $S^2$; more generally, head rotations in three dimensions ($3D$), specified by roll, pitch, yaw cover a large portion of a rotation group $\so(3)$ \cite{Anlk}. Similarly, the orientation-selective responses of neurons in the primary visual and auditory cortices, often give rise to spherical or cylindrical stimulus spaces \cite{Rng,Mzr,Brdf,Brsf1,Brsf2}. In contrast to previous models where metrics are externally imposed, our metrics arise intrinsically from the effective geometry of scale-free structure, making it both self-consistent and functionally grounded. Such a mapping requires topological consistency between the stimulus space and the boundary space of the embedding, which applies to all dimensions and scales. For practical considerations, we restrict our analysis to $1D$, $2D$ and $3D$ stimulus spaces, which cover a comprehensive set of experimentally observed cases.
% (with $d=1, 2, 3$) that capture $d$-dimensional orientation, 

\section{Results}
\label{sec:res}

\subsection*{Localized activity states capture the structure of stimulus space}
\begin{figure}[]
	%\begin{wrapfigure}{c}{0.75\textwidth}
	\centering
	\includegraphics[width=0.95\textwidth]{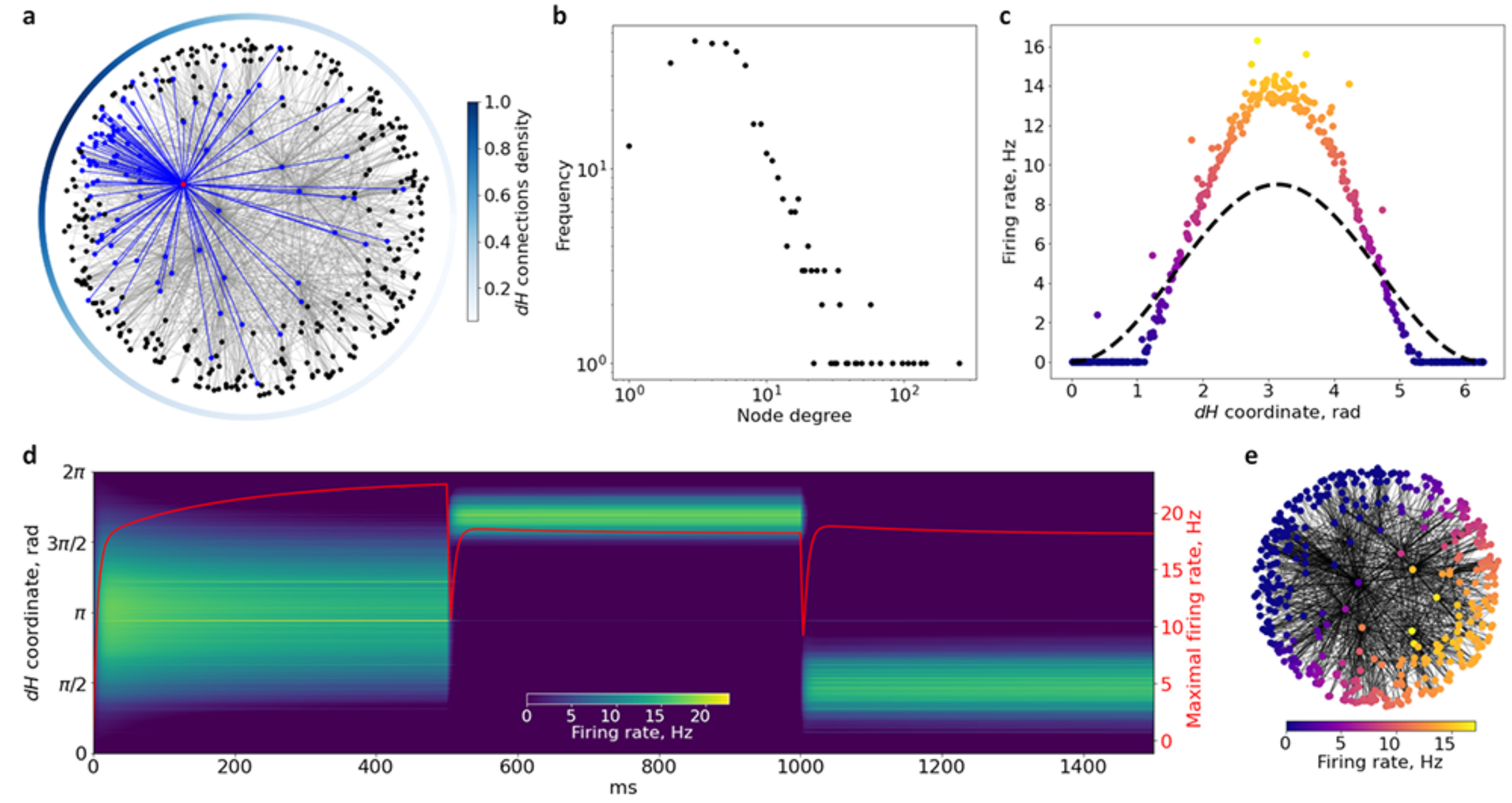}
	\caption{\footnotesize{ \textbf{Localized attractors over the stimulus spaces}. 
			\textbf{(a)} Hyperbolic network embedding into the Poincar\'e disk. The radial coordinate is associated with node degree. Color indicates the density of a neuron's neighbors along the boundary in a characteristic excitation profile.
			If the input distributes over the entire boundary, the activity shifts from the periphery to the bulk.
			\textbf{(b)} The effective negative curvature reflects the power-law distribution of node degrees.
			\textbf{(c)} The excitatory response to a stimulus field on the boundary (black dots). Response activity is localized on a small subset of neurons and suppressed on the periphery, induced solely by the network's effective geometry, without synaptic fine-tuning.
			\textbf{(d)} Network activity induced by stimuli of different widths and positions over $\sim$1.5 seconds. The amplitude of the bump attractor remains stable across stimulus sizes and locations (red line shows peak firing frequency).
			\textbf{(e)} Distribution of response activity across a scale-free network (cf. panel c).
	}}
	\label{f3:network_dynamics}
	%\end{wrapfigure}
\end{figure}

We began by investigating the emergence of localized activity states using rate-based models. We simulate the firing dynamics of the $N=400$ nodes, driven by excitation between neighbors according to the scale-free network structure without any additional fine-tuning of connection weights (Fig.~\ref{f3:network_dynamics}a, b). The inhibitory connections are random and homogeneous (see Method section for details). We started with a $1D$ stimuli space, \textit{i.e.}, neurons driven by inputs arranged along a topological circle, parameterized by a single angular coordinate inherited from network embedding (see Section \ref{sec:app}). 
%%%%%%%%%%%%%%%%%%%%%%%%%%%%%%%%%%%%%%%%%%%%%%%%%%%%%%%%%%%%%%%%%%%%%

%%%%%%%%%%%%%%%%%%%%%%%%%%%%%%%%%%%%%%%%%%%%%%%%%%%%%%%%%%%%%%%%%%%%%
%\noindent
First, when the network receives random, unstructured input, neuronal activity tends to concentrate within a dense cluster of highly connected nodes located near the center of the hyperbolic embedding. In contrast, under structured stimulation---localized within a compact angular domain--- we observe that the network's response shifts toward the periphery of the hyperbolic space, \textit{i.e.}, the embedding boundary, thereby mirroring the structure of the stimulus space. In other words, the activity in excitatory neurons acquires a limited angular span, \textit{i.e.}, formes a population activity bump localized along the angular coordinate (Fig.~\ref{f3:network_dynamics}c). Multiple longitudinal tests revealed that the attractor dynamic enforces the stability of the amplitude and localization of the activity bump regardless of the stimulus size and position. Fig.~ \ref{f3:network_dynamics}d shows the development of activity over time for cosine-shaped stimuli of different widths and localization along the angular coordinate. In other words, we observe an emergent orientational tuning, whose mechanism, however, differs from that used in the classical ring-shaped models \cite{Bsstt,SkgDir,StrngHD}. No adjustments of the synaptic connections or neighbor density alterations were used---this effect is of a purely geometro-topological nature and hence generic (Fig.~\ref{f3:network_dynamics}e).

The observed dynamics may correspond to a generic linearly organized firing, which may represent, \textit{e.g.},
a $1D$ place cell activity over a linear or circular track, or orientation-selective firing of head direction cells.  
%%%%%%%%%%%%%%%%%%%%%%%%%%%%%%%%%%%%%%%%%%%%%%%%%%%%%%%%%%%%%%%%%%%%%

\begin{figure}[H]
	%\begin{wrapfigure}{c}{0.75\textwidth}
	\centering
	\includegraphics[width=0.95\textwidth]{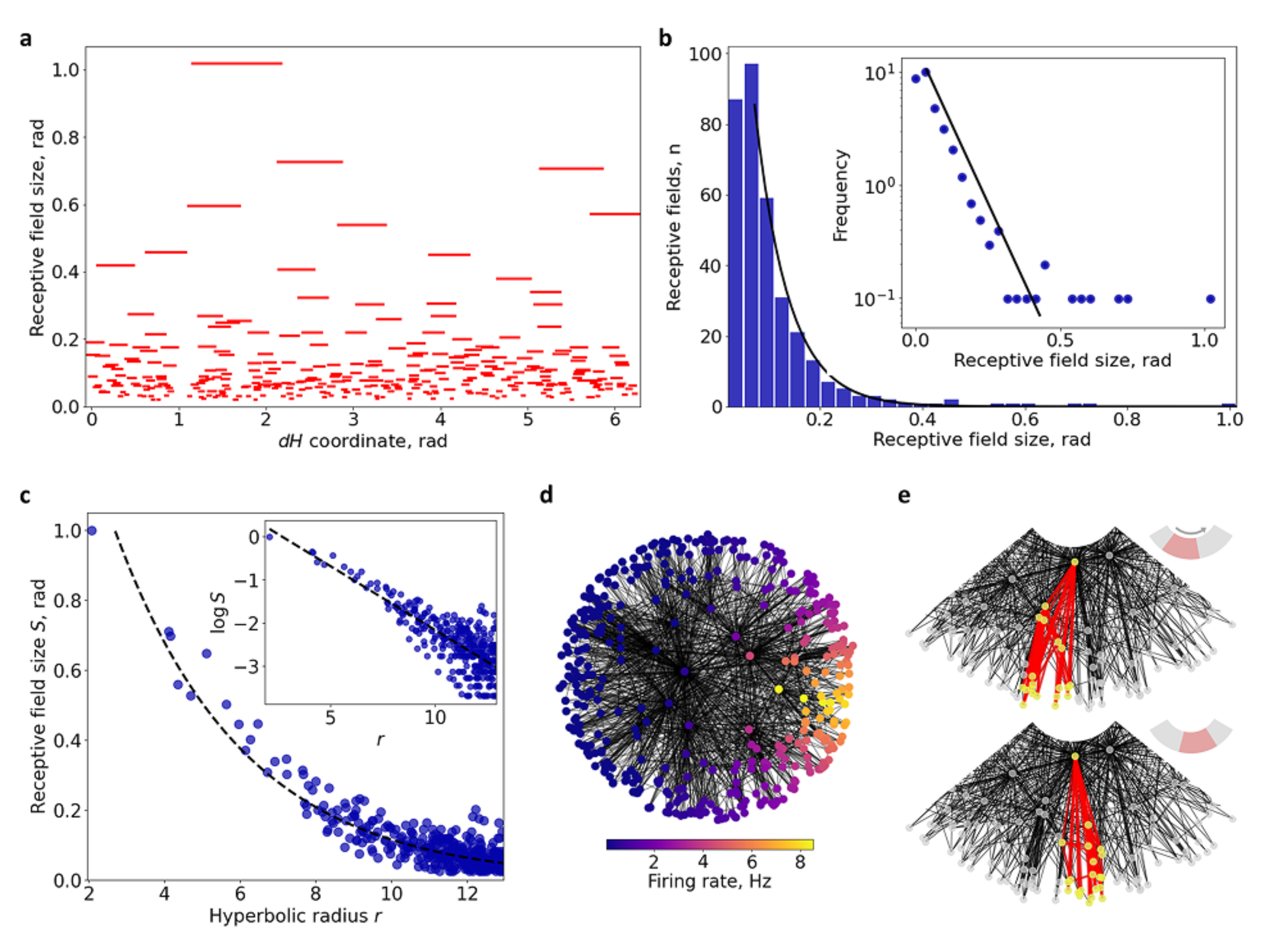}
	\caption{\footnotesize{
			\textbf{Hyperbolicity of Receptive Fields and Scale Sensitivity.}
			\textbf{(a)} The receptive fields of neurons in a hyperbolic network produce a complex tiling of the stimulus space.
			\textbf{(b)} The sizes of receptive fields in the frequency-based model correlate with the neurons' radial coordinates in the hyperbolic embedding.
			\textbf{(c)} The distribution of receptive field sizes follows an exponential law, consistent with experimental observations~\cite{zhang2023hippocampal}.
			\textbf{(d)} Localization of neural responses in a simulated hyperbolic network arises as a consequence of hyperbolicity---\textit{i.e.}, exponential concentration of cells toward the boundary.
			\textbf{(e)} A small stimulus displacement affects neurons with different receptive field scales differently: large receptive fields respond to broader shifts, while nested smaller fields exhibit greater sensitivity to fine changes.
	}}
	\label{f4:receptive_scaling}
	%\end{wrapfigure}
\end{figure}

%%%%%%%%%%%%%%%%%%%%%%%%%%%%%%%%%%%%%%%%%%%%%%%%%%%%%%%%%%%%%%%%%%%%%
The strengths of recurrent excitatory connections and peripheral inhibitions can vary within a wide but limited range without compromising structural flexibility, attractor robustness, and tractability. The system may lose attractor dynamics or become overexcited if the strength is too low or too high, typically outside physiological regimes. However, at regular connection strengths, the targeted attractor dynamics reliably emerges (see Methods, \ref{sec:met}). More detailed investigation of the stability conditions will be provided elsewhere and is beyond the scope of this study.

\subsection*{Network Model and stimulus space embedding approach}
Next, we study the sizes of the receptive fields and their layouts by identifying the subdomains of the stimulus space that elicit reliable responses from individual neurons. Following the standard experimental methodology, we identified each field as a domain in the boundary space to which a given neuron responded with confidence, \textit{i.e.}, above a certain threshold. Computationally, we swept a narrow stimulus along the boundary, slowly enough to saturate the amplitude of the bump attractor in each position (see Methods, \ref{sec:met}).
%%%%%%%%%%%%%%%%%%%%%%
%%%% NEW

The resulting receptive fields form a dense coverage of the stimulus space (Fig.~\ref{f4:receptive_scaling}a), and, more importantly, exhibit exponential nesting of observed sizes(Fig.~\ref{f4:receptive_scaling}b), which is consistent with the experimentally observed distribution of place field sizes \cite{Rch,Elv}. 

Specifically, the size of a receptive field, $S$, and the corresponding neuron's radial coordinate, $r$ (Fig.~\ref{f4:receptive_scaling} c), are related as
\begin{equation}
	S \propto e^{-\alpha r},
	\label{eq:r_vs_s}
\end{equation}
and implies a direct relationship between a neuron's receptive field size $S$ and its connectivity degree $k$,
\begin{equation}
	S \propto k^{-\beta}.
	\label{eq:s_self}
\end{equation}
Here $\alpha,\beta>0$ are model–dependent constants; in particular, the precise numerical value of $\beta$ is controlled both by parameters of the network model and neural dynamic model, and a detailed estimation lies beyond the scope of the present study.
\noindent

%%%%%%%%%%%%%%%%%%%%%%%%%%%%%
The larger the receptive field of a neuron, the less its activity varies in response to small changes in stimulus. For each particular stimulus, both large and small receptive fields are engaged in driving an ensemble of active cells, descending from the upscale to the smallest ones (Fig.~\ref{f4:receptive_scaling}d). Consequently, the localized activity ensemble comprises neurons that span all hierarchical scales along the radial axis of the embedding. As expected, a small change in stimulus strongly affects cells with smaller receptive fields, whereas the activity of neurons with large receptive fields does not alter significantly, \textit{i.e.}, have different scale sensitivity (Fig.~\ref{f4:receptive_scaling}e). Thus, the activity of cells with large receptive fields provides a ``context" for the active cells whose receptive fields are nested within them, in a sense uniting the nested cells on a larger scale \cite{Wlns}.

\subsection*{Spiking dynamics in hyperbolic network}

Spiking dynamics in hyperbolic networks also produces localized attractor dynamics. We simulated a population of $400$ excitatory and $400$ inhibitory spiking neurons, using Izhikevich's model ---a choice motivated by its numerical effectiveness and ability to produce a comprehensive scope of physiologically realistic dynamic behaviors \cite{izhikevich2003}. As in the rate-based model, excitatory connections were wired into a scale-free network, while inhibitory connections were established randomly (see Methods, \ref{sec:met}). Synaptic strengths were tuned to match the effective receptive field sizes observed in the corresponding rate-based network under identical stimulation. The external inputs were applied to the embedding $1D$ boundary , with the addition of random noise to support initial excitation.
%%%%%%%%%%%%%%%%%%%%%%%%%%%%%%%%%%%%%%%%%%%%%%%%%%%%%%%%%%%%%%%%%%%%%

\begin{figure}[H]
	%\begin{wrapfigure}{c}{0.75\textwidth}
	\centering
	\includegraphics[width=0.98\textwidth]{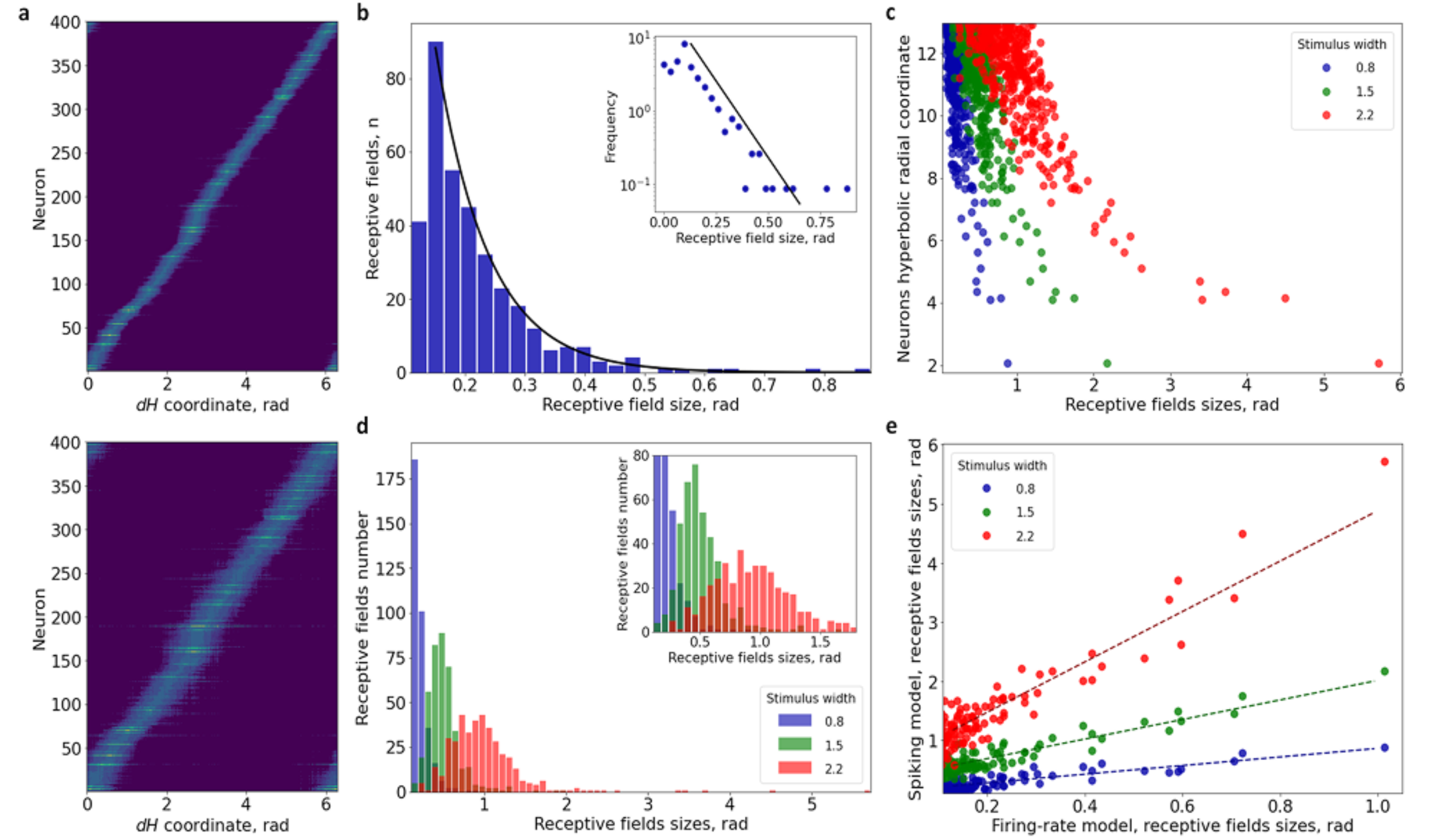}
	\caption{\footnotesize{
			\textbf{Place field scaling.} Different widths and receptive field sizes in the firing-rate model.
			\textbf{(a)} Distribution of neuron activity across the stimulus space in the spiking model for both narrow and wide stimuli.
			\textbf{(b)} Distribution of receptive field sizes for a narrow stimulus.
			\textbf{(c)} Dependence of receptive field size on the neuron's radial coordinate for stimuli of different widths.
			\textbf{(d)} Shift in the distribution of receptive fields with increasing characteristic stimulus width.
			\textbf{(e)} Ratio of receptive field sizes between the spiking and rate-based models (with the narrowest stimulus width fixed in the rate model).
	}}
	\label{f5:placefield_scaling}
	%\end{wrapfigure}
\end{figure}
\vspace{-\baselineskip}

%%%%%%%%%%%%%%%%%%%%%%%%%%%%%%%%%%%%%%%%%%%%%%%%%%%%%%%%%%%%%%%%%%%%% %\noindent
As anticipated, the angular domains of the resulting neuronal responses covered the entire stimulus space and distributed exponentially in size (Fig.~\ref{f5:placefield_scaling}a,b), according to each neuron's radial coordinate in hyperbolic embedding (Fig.~\ref{f5:placefield_scaling}c). Fig.~\ref{f5:placefield_scaling}a shows the distribution of spike events along the boundary coordinate associated with the external stimulus space, as in the firing-rate model. Thus, the spiking network produces a direct analogue of the rate-based model.

Next, we studied how the structure of the place field depends on the size of the environment. Experiments demonstrate that in small enclosures (a couple of meters in length), place fields are typically compact, with sizes distributing normally around a certain typical value \cite{scleidorovich, mainali,Rch, GevaSagiv2015, Moser2014}, while environments that significantly exceed the animals' size (\textit{for example}, rats over $50$-meter long track \cite{Rch}, or flying bats in large open environments \cite{Elv}), place fields are multiply connected and nested exponentially (Fig.~\ref{f1:receptive_fields}d). 

We tested whether the receptive fields respond to the relative scale of the stimulus and found that, indeed, the characteristic size of the receptive fields increases with stimulus width (Fig.~\ref{f5:placefield_scaling}d). In larger environments, which we simulated by confining stimuli to smaller domain on the hyperbolic boundary, the result is similar to experimental observations for place cells. With wider stimuli (associated with relatively small environments), the receptive field sizes for the neurons with a large radial coordinate distribute more symmetrically---notice the appearance of a left tail on the histogram (Fig.~\ref{f5:placefield_scaling}d). Outside of a thin right tail that accounts for the rare large receptive fields, the distribution becomes similar to the normal. This shift is consistent with experimental findings, where rats in a large environment (\textit{, i.e.}, a relatively small scale of stimulus to the size of the environment) have a close-to-exponential distribution of receptive field sizes, but exhibit fields with typical sizes in a small space, such as a laboratory cage \cite{zhang2023hippocampal}.

Curiously, the receptive field sizes in both spiking and the rate-based networks with identical connectivity, computed for the same neurons (located at the same nodes), are related approximately linearly (Fig.~\ref{f5:placefield_scaling}e). The two receptive field maps have the same overall structure and the same granularity. This correspondence suggests that receptive fields form at the populational level and that models based on mean-field and effective synaptic couplings are sufficient to capture the essential properties of receptive fields' structure and stimulus encoding.

\subsection*{Validating Hyperbolicity with Rat CA1 Recordings}

\begin{figure}[H]
	%\begin{wrapfigure}{c}{0.75\textwidth}
	\centering
	\includegraphics[width=0.98\textwidth]{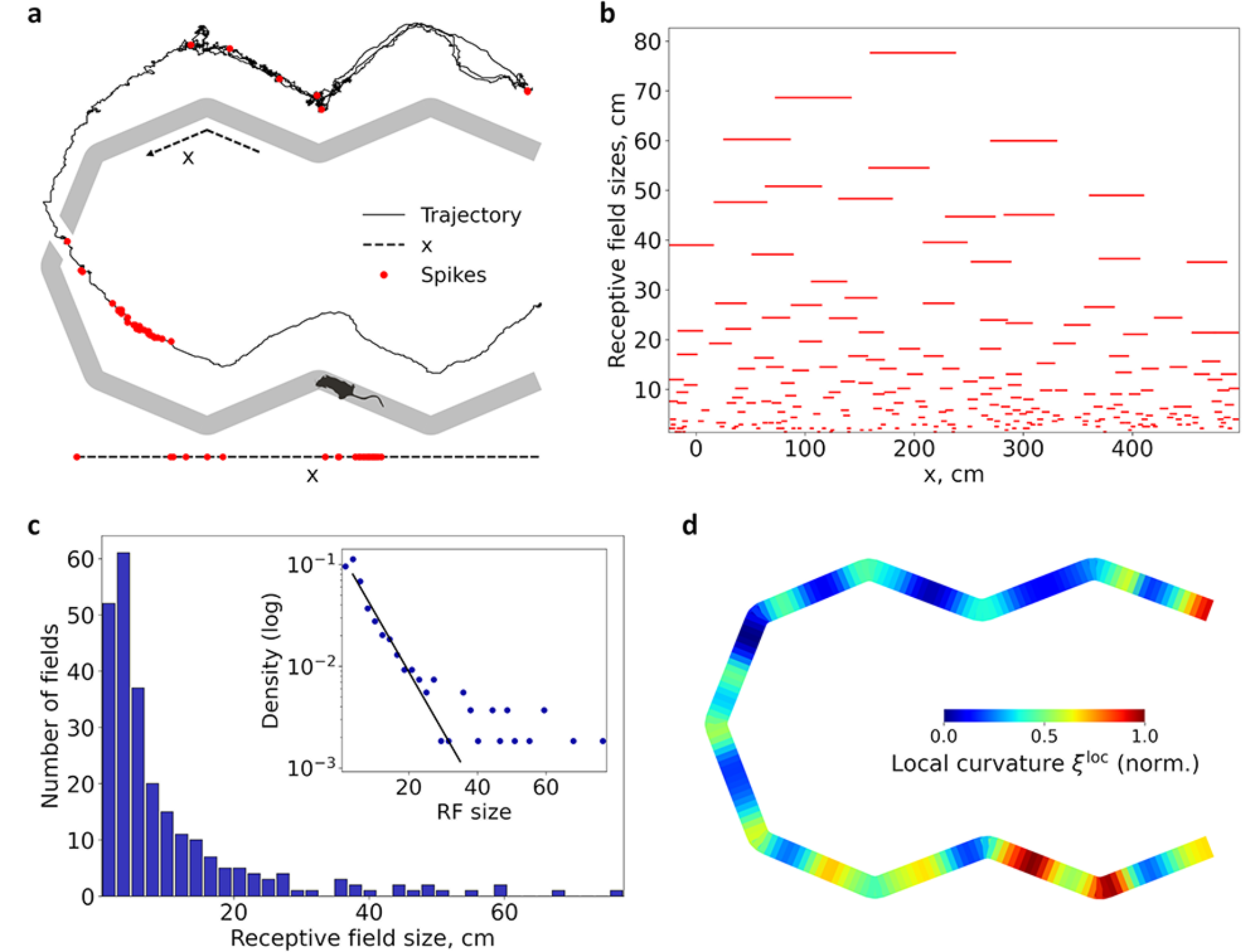}
	\caption{\footnotesize{
			\textbf{Experimental verification of the place fields’ statistics on a linear track}.
			\textbf{(a)} Schematic representation of the track’s runway (gray stripe, top-down view) and a segment of the rats trajectory (solid black line, lifted for visibility). The dashed black line at the bottom indicates the linearized coordinate of the track (compressed scale), and the red dots correspond to spikes of a neuron. 
			\textbf{(b)} The observed net coverage of the track by the place fields align with the results reported previously \cite{Rch,zhang2023hippocampal}. 
			\textbf{(c)} Distribution of receptive field sizes; black line in the inset denotes the exponential fit to the distribution. 
			\textbf{(d)} Variations of local curvature $\xi^{\mathrm{loc}}$ (unit interval normalization). The maxima of $\xi^{\mathrm{loc}}$ are concentrated at the junctions of the composite track segments, which attract the animal’s attention and are consistent with natural exploratory behavior.
	}}
	\label{f6:rf_hyperbolicity}
	%\end{wrapfigure}
\end{figure}

To test the outcomes of the model, we analyzed the activity of place cells recorded in a rat exploring a linear path. The track was composed of ten sections, as illustrated in Fig.~\ref{f6:rf_hyperbolicity}a, which were slowly deformed relative to the 2D ambient environment, without inducing a global remapping.
%%%%%%%%%%%%%%%%%%%%%%%%%%%%%%%%%%%%%%%%%%%%%%%%%%%%%%%%%%%%%%%%%%%%%

%%%%%%%%%%%%%%%%%%%%%%%%%%%%%%%%%%%%%%%%%%%%%%%%%%%%%%%%%%%
\noindent Previous analyses of these data have shown that place fields remain stable in the linear reference frame, which allows us to focus on their linear sizes \cite{eLife}. As shown on Fig.~\ref{f6:rf_hyperbolicity}b, place fields vary in size and their distribution is well fit by an exponential profile (Fig.~\ref{f6:rf_hyperbolicity}c), \textit{i.e.}, indeed exhibits a hyperbolic behavior. 
We further investigated the local curvature of the coverage, $\xi^{\mathrm{loc}}$, estimating from the corresponding density of receptive field sizes along the $x$-coordinate (see Methods, \ref{sec:met}):
\bigskip
\begin{equation*}
	p(s) = \frac{\xi \sinh\big(\xi(s_{\max}-s)\big)}{\cosh(\xi s_{\max})-1},
\end{equation*}
%\bigskip
where $p(s)$ is the probability density of place field sizes and $s_{\max}$ is the maximal field size. Fig.~\ref{f6:rf_hyperbolicity}f shows the variation of local curvature $\xi^{\mathrm{loc}}$ across the track. 

As can be seen (Fig.~\ref{f6:rf_hyperbolicity}d), pronounced peaks of $\xi^{\mathrm{loc}}$ appear at the junctions between track segments. These junctions attract the animal’s attention, consistent with natural exploratory behavior, and we thus observe a deepening of the hyperbolic representation in these regions despite the overall moderate curvature of the coverage, in agreement with previous results in large-scale environment \cite{Rch}.
%%%%%%%%%%%%%%%%%%%%%%%%%%%%%%%%%%%%%%%%%%%%%%%%%%%%%%%%%%%%%%%%%%%

\begin{figure}[h]
	%\begin{wrapfigure}{c}{0.75\textwidth}
	\centering
	\includegraphics[width=0.9\textwidth]{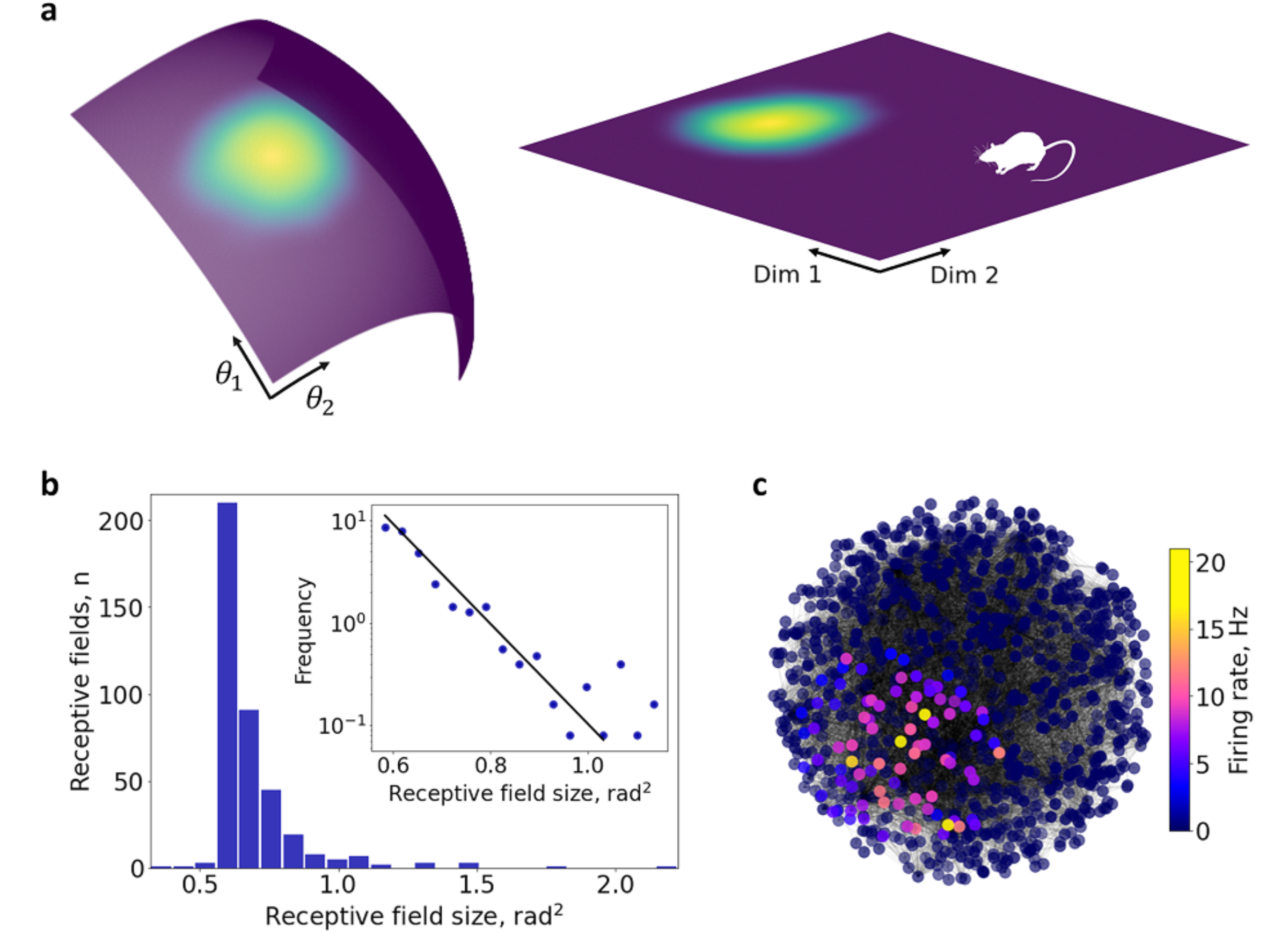}
	\caption{\footnotesize{\textbf{$2D$ stimulus space and planar place fields.}
			\textbf{(a)} A hyperbolic network embedding with a $2D$ boundary is associated with the $2D$ stimulus space, illustrating the correspondence between receptive fields observed on the embedding boundary and the spatial receptive fields of place cells.
			\textbf{(b)} Localized activity along the $2D$ boundary of a $3D$ hyperbolic network embedding. 
	}}
	\label{f7:2Dembedding}
	%\end{wrapfigure}
\end{figure}

%%%%%%%%%%%%%%%%%%%%%%%%%%%%%%%%%%%%%%%%%%%%%%%%%%%%%%%%%%%%%%%%%%%
The length of the track, around $5$ meters, being intermediate in scale between a small environment such as a laboratory cage and a large-scale open environment, provides an opportunity to obtain place fields of different sizes and to observe the emergence of low curvature hyperbolicity. Indeed, the characteristic curvature is substantially smaller ($n=253$, $\xi^{\mathrm{glb}}=-0.14$; $n=167$, $\xi^{\mathrm{glb}}=-0.13$; $n=182$, $\xi^{\mathrm{glb}}=-0.09$) than that observed in large-scale environments \cite{Rch}.

\subsection*{Higher dimensions}
The model also reproduces physiologically viable receptive fields in hyperbolic networks embedded in higher-dimensional hyperbolic spaces---such as $3D$- or $4D$ hyperbolic 'Poincar\'e balls' with their respective boundaries. By associating these boundaries with higher-dimensional stimulus spaces, the model generates multidimensional receptive fields that occupy planar or volumetric physical spaces \cite{Elv,Yrts,Jffr,Hymn}. We simulated such fields using equal numbers of excitatory and inhibitory neurons, scaling the overall population size to match the requirements of $2D$ and $3D$ spaces (see Methods, \ref{sec:met}).
%%%%%%%%%%%%%%%%%%%%%%%%%%%%%%%%%%%%%%%%%%%%%%%%%%%%%%%%%%%%%%%%%%%%%

\begin{figure}[H]
	\centering
	\includegraphics[width=0.9\textwidth]{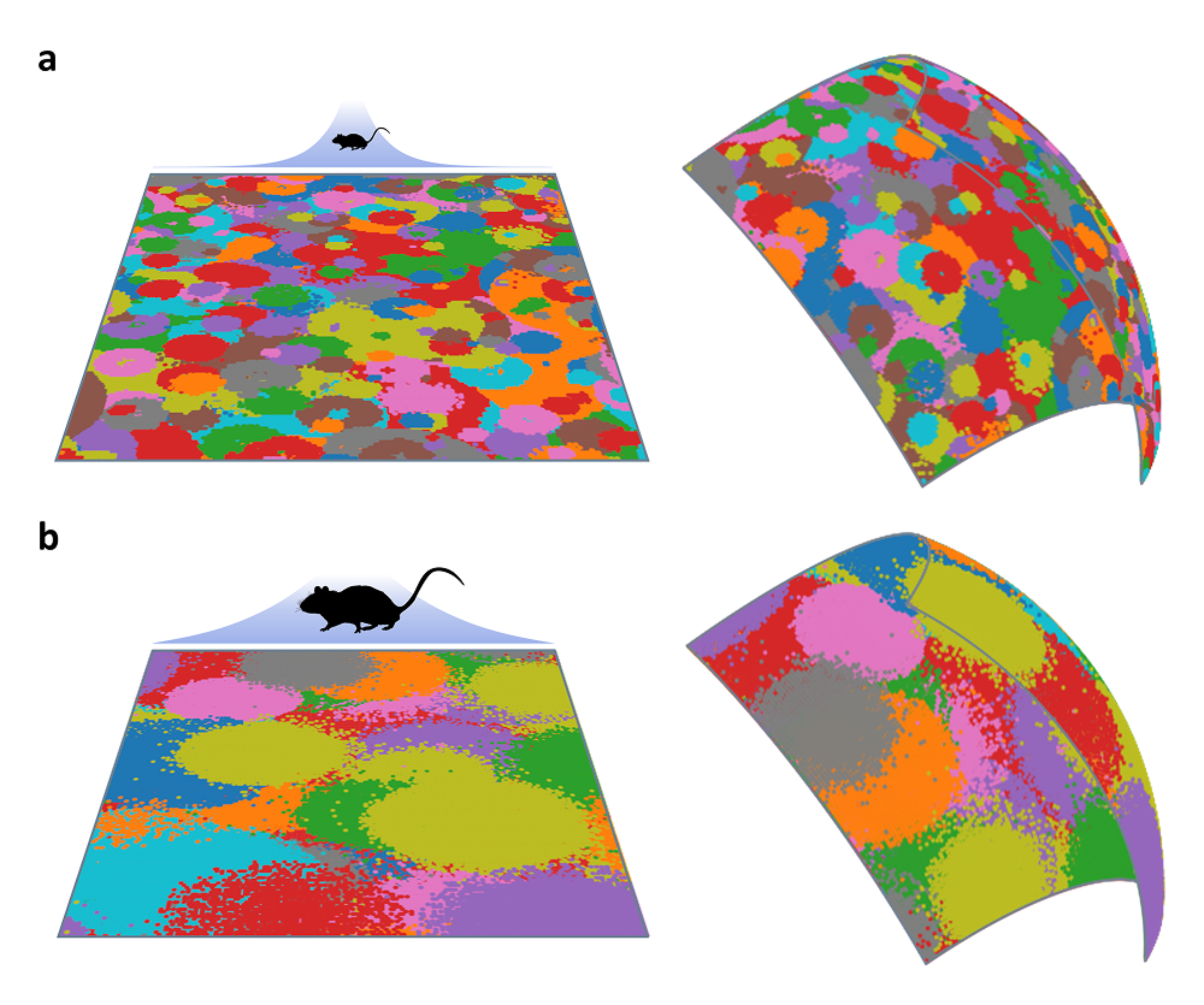}
	\caption{\footnotesize{\textbf{Changing the scale of the stimulus relative to the size of the environment alters the receptive fields.}
			The colored dots show the $2D$ place fields produced by the same network under different stimulus sizes, corresponding to larger environment (\textbf{a}) and smaller environment (\textbf{b}).
	}}
	\label{f8:stimulus_scale_effect}
\end{figure}

%%%%%%%%%%%%%%%%%%%%%%%%%%%%%%%%%%%%%%%%%%%%%%%%%%%%%%%%%%%%%%%%%%%%%
The place field map produced in $2D$ case (Fig.~\ref{f7:2Dembedding}a), the receptive field sizes follow an exponential distribution, as in the $1D$ case (Fig.~\ref{f7:2Dembedding}b). The activity domain remains localized along the embedding boundary, showing suppression of peripheral activity and sharpened stimulus representation (Fig.~\ref{f7:2Dembedding}c). The characteristic sizes of the receptive fields, as in the case of a $1D$ stimulus space, depend on the scale of the stimulus relative to the environment. Fig.~\ref{f8:stimulus_scale_effect} schematically shows the changes in the layout of $2D$ receptive fields for different stimulus sizes.

\begin{figure}[H]
	\centering
	\includegraphics[width=0.98\textwidth]{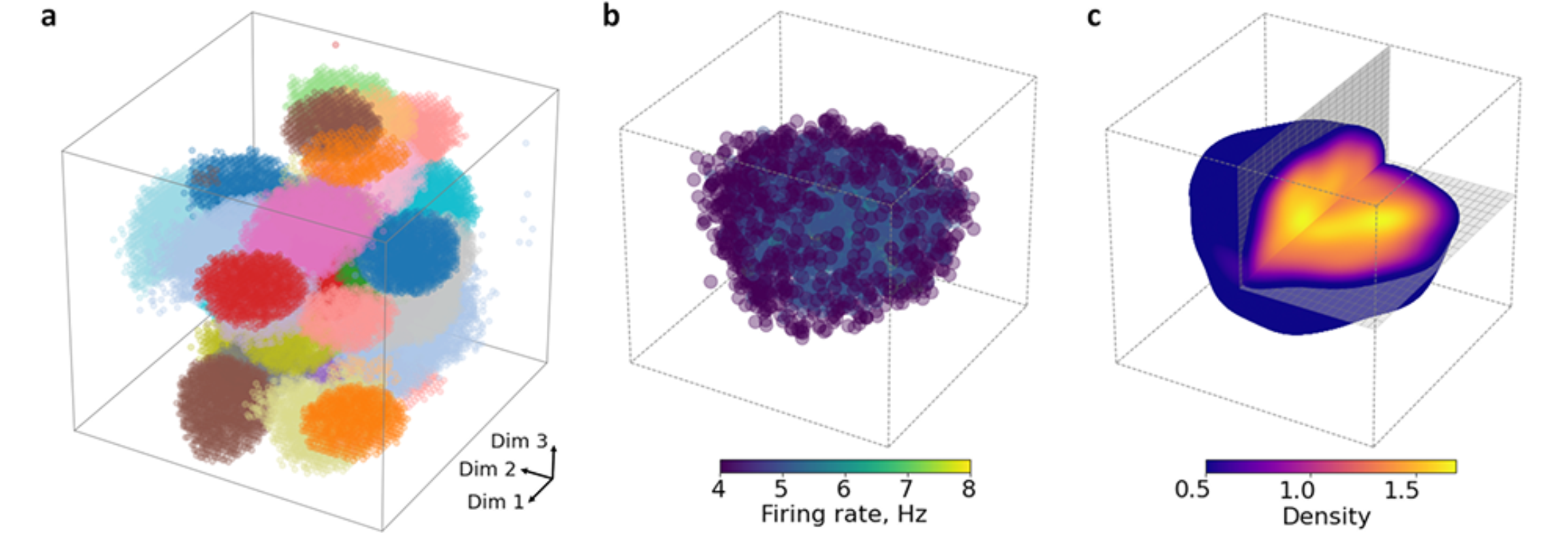}
	\caption{\footnotesize{
			\textbf{$3D$ receptive fields.}
			\textbf{(a)} Receptive fields of several randomly selected neurons, projected from the $3D$ boundary of the network embedding into Euclidean $3D$ space (each shown in a different color).
			\textbf{(b)} Receptive field of a single neuron; each point represents a stimulus position that elicited a stable firing response of at least 4~Hz (color indicates observed firing rate).
			\textbf{(c)} Receptive field of the same neuron as in panel (b), shown as a Gaussian kernel estimate of activation strength over the stimulus configuration space. Only regions with kernel density above 0.5 are shown. A cut-out section illustrates the internal depth structure of the $3D$ receptive field.
	}}
	\label{f9:3Dreceptive_fields}
\end{figure}

In the case of a $3D$ stimulus space, the corresponding $4D$ network embedding cannot be directly visualized. However, the model's principles and implementation do not change: we observe the formation of distinct receptive fields for individual neurons, which collectively tile the $3D$ volume when the embedding boundary is projected onto it (Fig.~\ref{f9:3Dreceptive_fields}a). Each receptive field exhibits a pronounced increase in activation frequency toward its center, as shown in the environment depicted in (Fig.~\ref{f9:3Dreceptive_fields}b,c). Thus, the emergence of receptive fields from the structure of scale-free networks is a generic property of the model.

\section{Discussion}
\label{sec:dis}

Neural dynamics across diverse brain networks is shaped by how individual neurons' receptive fields partition the stimulus space \cite{Ptl}. This organizational principle plays a central role in linking the structure of the stimulus space---whether sensory, spatial, or conceptual---to neuronal dynamics and may represent a fundamental mechanism of information encoding. The physiological mechanism underlying this phenomenon remained puzzling: how neural dynamics become functionally coupled to external stimulus spaces, and which properties of network connectivity support such mappings are questions that demand fundamental understanding.
In some cases, the desired dynamics can be achieved by fine-tuning synaptic strengths according to the proximity of receptive fields and aligning network topology with that of a target manifold, as, \textit{e.g.}, in attractor network models of angle-selective neuronal firing. However, in deep brain networks, where inputs are highly processed and cannot be naively geometrized, the link between the synaptic architecture and the internal dynamics becomes obscured. A biologically plausible mechanism for generating receptive fields in such networks is therefore more likely to rely on a different organizing principle---one that is self-contained, more general, and independent of externally observed phenomenology.
%%%%%%%%%%%%%%%%%%%%%%%%%%%%%%%%%%%%%%%%%%%%%%%%%%%%%%%%%%%

%%%%%%%%%%%%%%%%%%%%%%%%%%%%%%%%%%%%%%%%%%%%%%%%%%%%%%%%%%%
The proposed approach allows capturing the structure of stimulus space through network dynamics, via a mechanism modulated---or even induced---by the network's connectivity. The model relies entirely on the network's endogenous topological structure, rather than the design of synaptic efficacies based on external assessments. In particular, scale-free networks allow associating the stimulus space with their effective hyperbolic boundaries. This connection requires no assumptions about the physical arrangement of neurons---the embedding is strictly functional---and offers a natural, unifying explanation for several experimental observations, such as the appearance of low-dimensional receptive fields, their layout, the empirically observed distributions of their shapes and sizes, their network-level origin, and so forth. A curious and experimentally testable prediction of the model is that the scale of a receptive field is related to the connectivity degree of the corresponding neuron.

Importantly, the scale-free property does not require locally specified connectivity; rather, it arises from a more general statistical property---the hierarchical structure of connectivity, characterized by the distribution of node degrees across the network. As long as the global degree hierarchy is maintained, the network preserves its functional properties under changes in local connectivity, synaptic plasticity, and alterations of external input. Within this framework, the stimulus space may, in principle, have arbitrary dimensionality, provided a consistent functional mapping onto the network's effective boundary. However, the finite size of biological neural networks may impose constraints on the range of receptive field dimensionalities---a consideration that lies beyond the scope of this discussion.

In this study, we maintained the functional embedding curvature close to constant \cite{krioukov2010hyperbolic,Aldc}. Natural variability of synaptic connectivity may produce higher-order effects, \textit{ e.g.}, fluctuations of the local embedding curvatures and maximal hyperbolic embedding radii, thereby affecting the resulting neural dynamics. Experimental studies have shown that receptive fields undergo tuning and detuning across multiple timescales---an effect that may reflect a progressive deepening of the effective hyperbolic embedding in response to the intensity and duration of exploration, as well as other behavioral factors \cite{zhang2023hippocampal}. Our approach offers a framework for investigating such structural variability and its functional implications.

Lastly, localized populational excitation states---bump attractors---are robust to noise at large timescales and align with the slow kinetics of external world dynamics. Such states support population-level information processing, connecting rapidly fluctuating neuronal responses with slower, mean-field ensemble dynamics. This allows a localized ensemble of excited neurons to encode the current stimulus state through a functional partition of the stimulus space with receptive fields. On the other hand, the emerging internal representations of the stimulus are also directly shaped by the receptive fields’ layout. If receptive fields are hyperbolically nested, as observed experimentally and reproduced in our model, the excited ensemble simultaneously includes neurons at multiple scales, each exhibiting different sensitivity to stimulus variations. The dynamics of neuronal responses effectively inherit the scale-free property, exhibiting sensitivity to perturbations across all spatial scales of the stimulus space. Although the role of this organization in information processing is yet to be clarified, our approach provides a means to explore it via structure-function relationships of receptive fields organization.

The observation that our embedding-based framework for stimulus space, grounded in the effective geometry of scale-free networks, captures several observed phenomena suggests that it may provide a useful model for understanding the activity in brain networks, including the hippocampus.

\textbf{Acknowledgements}
%\label{sec:ack}
We express our gratitude to Dr. Anton Chizhov for fruitful discussions.

\newpage
\section{Methods}
\label{sec:met}

\subsection*{Network Model and stimulus space embedding approach}
We employ the approach of Krioukov \textit{et al.}~\cite{krioukov2010hyperbolic,Aldc}, which allows one to generate random scale-free graphs $G(V,E)$ with controllable parameters, including the power-law degree distribution exponent $ \gamma$. The network nodes are embedded in hyperbolic space $ \mathbb{H}^{d}$ (with $ d \geq 2$), where each node is assigned one hyperbolic radial coordinate $r$ and $d-1$ angular coordinates $ \boldsymbol{\theta}^{\mathrm{emb}}=(\theta^{\mathrm{emb}}_1, \theta^{\mathrm{emb}}_2, \dots, \theta^{\mathrm{emb}}_{d-1})$, reflecting an effective geometric structure.
For simplicity, we consider stimulus spaces that define an $ d^{\mathrm{stim}}$-dimensional orientation (sphere  $ \mathbb{S}^{d^{\mathrm{stim}}}$). A stimulus is defined as a scalar field over $ \mathbb{S}^{d^{\mathrm{stim}}}$, for example, exhibiting a cosine-like profile in the stimulus coordinates $ \boldsymbol{\theta}^{\mathrm{stim}}=(\theta_1^{\mathrm{stim}}, \theta_2^{\mathrm{stim}}, \dots, \theta_{d^{\mathrm{stim}}}^{\mathrm{stim}})$. In both rate-based and spiking models of neural activity (see the next subsection), the $ d^{\mathrm{stim}}$-dimensional stimulus space is associated with the $d-1$ -dimensional embedding boundary via the mapping $ \boldsymbol{\theta}^{\mathrm{emb}} \mapsto \boldsymbol{\theta}^{\mathrm{stim}}$, under the dimensionality-matching condition $ d^{\mathrm{stim}}=d-1$.

\vspace{-10pt}
\subsection*{Spiking Neuron Model}
To simulate spiking dynamics, we used the Izhikevich neuron model due to its numerical efficiency and its ability to reproduce a wide range of physiologically realistic spiking and bursting behaviors with low computational cost~\cite{izhikevich2003,IzhPh2}. The model is defined by the following set of differential equations.
\begin{equation*}
	\begin{aligned}
		\frac{dv}{dt} &= 0.04v^2 + 5v + 140 - u + I(t), \\
		\frac{du}{dt} &= a(bv - u),
	\end{aligned}
\end{equation*}
with the auxiliary after-spike resetting condition:
\begin{equation*}
	\text{if } v \geq 30 \text{ mV, then }
	\begin{cases}
		v \leftarrow c, \\
		u \leftarrow u + d.
	\end{cases}
\end{equation*}
Here, $ v(t)$ is the membrane potential, $ u(t)$ is the membrane recovery variable, $ I(t)$ is the input current, and $ a, b, c, d$ are model parameters that determine the neuron's dynamical class (\textit{e.g.}, regular spiking, fast spiking, or bursting). Following the classical configuration introduced by Izhikevich~\cite{izhikevich2003}, we assigned neuron parameters to reproduce realistic heterogeneity within excitatory and inhibitory populations. For excitatory neurons, parameters were set as $ a=0.02$, $ b=0.2$, $ c=-65 + 15r_e^2$, and $ d=8 - 6r_e^2$, where $ r_e \sim \mathcal{U}(0, 1)$ is a uniformly distributed random variable. This configuration spans a range from regular spiking to chattering behavior. For inhibitory neurons, the parameters were $ a=0.02 + 0.08r_i$, $ b=0.25 - 0.05r_i$, $ c=-65$, and $ d=2$, with $ r_i \sim \mathcal{U}(0, 1)$, capturing a spectrum from fast-spiking to low-threshold spiking interneurons.

The excitatory population connections follow the adjacency matrix of the scale-free graph generated, modulated by a synaptic conductance parameter $ g_e=3$. Interactions within the inhibitory population and between inhibitory and excitatory populations are defined by random connectivity matrices, scaled by conductance parameters $ g_i=4$, $ g_{ei}=1.5$, and $ g_{ie}=1.5$, respectively. The synaptic weight matrix $ S \in \mathbb{R}^{N \times N}$ determines the total interactions of the model, including structured random and scale-free components.

Neurons receive an input current composed of recurrent synaptic input and external stimulation. The total input to the neuron $ i$ at time $ t$ is defined as 
$ I_i(t)=I^{\mathrm{stim}}_i(t) + \sum_{j \in \mathcal{F}(t)} S_{ij}$, 
where $ \mathcal{F}(t)$ is the set of indices of neurons that were activated at time $ t$, and $ S_{ij}$ is the synaptic weight of the neuron $ j$ to neuron $ i$. The input of external stimuli for each neuron is defined as 
\begin{equation*}
	I^{\mathrm{stim}}_i(t) =
	\begin{cases}
		g^{\mathrm{stim}} \cdot f^{\mathrm{stim}}_i(t) + g^{e,\mathrm{noise}} \cdot \xi_i(t), & \text{if neuron } i \in \text{ excitatory population}, \\
		g^{i,\mathrm{noise}} \cdot \xi_i(t), & \text{if neuron } i \in \text{ inhibitory population}
	\end{cases}
\end{equation*} 
where $ f^{\mathrm{stim}}_i(t)$ is the value of the stimulus field in the receptive coordinates of neuron $ i$, and $ \xi_i(t) \sim \mathcal{N}(0,1)$ is Gaussian white noise. The parameters $ g^{\mathrm{stim}}=5$, $ g^{e,\mathrm{noise}}=1$, and $ g^{i,\mathrm{noise}}=4$ scale the contributions of stimulus and noise to excitatory and inhibitory neurons, respectively.

As a stimulus field for detecting receptive fields in the spiking model, we use a narrow profile that smoothly decays toward the periphery, defined as  $ I^{\mathrm{stim}}(\boldsymbol{\theta})=\exp\left( -\|\boldsymbol{\theta}_{\mathrm{stim}} - \boldsymbol{\theta}\| / w_{\mathrm{stim}} \right)$, 
where $ \boldsymbol{\theta}_{\mathrm{stim}}$ denotes the current peak location on the stimulus sphere, and $ w_{\mathrm{stim}}$ controls the spatial width of the stimulus.
\vspace{-10pt}

\subsection*{Population Rate Model with Synaptic Depression}
We used a classical rate-based model, in which the dependence of the firing rate on synaptic current is represented by a threshold-linear function, extended to incorporate synaptic depression. The dynamics of the synaptic current follow: 
\begin{equation*}
	\tau \frac{dI(t)}{dt}=-I(t) + I^{\mathrm{stim}}(t) + I^{\mathrm{rec}}(t),
\end{equation*} 
where $ \tau$ is the time constant of excitatory transmission. The recurrent current $ I^{\mathrm{rec}}(t)$ represents the input of internal connections within the population and is defined as 
\begin{equation*}
	I^{\mathrm{rec}}_i(t)=\sum_j S_{ij} \cdot P_{\mathrm{rel}}(t) \cdot r(t),
\end{equation*} 
where $ S \in \mathbb{R}^{N \times N}$ is the internal connectivity matrix that determines how the nodes influence each other. Excitatory connections are defined according to the adjacency matrix of a scale-free graph generated and are scaled by excitatory conductance $ g^e$. All remaining entries in the matrix, that is, those that do not correspond to excitatory connections, represent inhibitory influences and are randomly sampled from a uniform distribution in the interval $[-1, 0]$, then scaled by inhibitory conductance $ g^i$. In this way, the matrix $ S$ simultaneously encodes structured excitatory input and homogeneous random inhibitory connectivity.
Synaptic depression modifies transmission through the probability of release $ P_{\mathrm{rel}}(t)$ ~\cite{rossum}, which evolves as 
\begin{equation*}
	\tau_{\mathrm{depr}} \frac{dP_{\mathrm{rel}}(t)}{dt}=P_0 - \left[1 + \tau_{\mathrm{depr}} r(t)(1 - f)\right] P_{\mathrm{rel}}(t),
\end{equation*} 
where $ \tau_{\mathrm{depr}}=500$ ms is the recovery time constant, $ P_0=1$ is the baseline release probability, and $ f=0.8$ is the depression factor. The depression factor $ f=0.8$ describes how much the synapses depress after each spike, effectively modifying the release probability via $ P_{\mathrm{rel}} \rightarrow f P_{\mathrm{rel}}$.
The firing rate is determined by a simple threshold function $r(t)=[I(t)]_+,$ where $ [\cdot]_+$ denotes rectification, corresponding to taking the non-negative part: $ [x]_+=\max(x, 0)$.

To investigate the localization of population activity that resembles a bump attractor, we used a cosine-shaped input profile defined as $ I^{\mathrm{stim}}(\theta)=g_{\mathrm{stim}}^{\mathrm{fr}} \cdot \left[\cos(2\pi(\theta - \theta_{\mathrm{stim}})) + 1\right]$, where $ \theta_{\mathrm{stim}}$ indicates the center of stimulation and $ g_{\mathrm{stim}}^{\mathrm{fr}}=9$ controls the amplitude. For receptive field analysis, as in the spiking model, we used a narrow profile that smoothly decays toward the periphery, defined as $ I^{\mathrm{stim}}(\boldsymbol{\theta})=g_{\mathrm{stim}}^{\mathrm{fr}} \cdot \exp(-\|\boldsymbol{\theta}_{\mathrm{stim}} - \boldsymbol{\theta}\| / w_{\mathrm{stim}})$, where $ \boldsymbol{\theta}_{\mathrm{stim}}$ denotes the current peak location in the stimulus sphere, and $ w_{\mathrm{stim}}$ controls the spatial width of the stimulus, and $ g_{\mathrm{stim}}^{\mathrm{fr}}=9$.

\subsection*{Experimental Recordings}
A full description of the methodology is provided in ~\cite{eLife}. The CA1 hippocampal activity was recorded in rats navigating a linear track with a total length of approximately $5$ meters, composed of ten $\sim50$ cm sections. These segments were arranged into a flexible U-shaped track. The two arms of the track moved independently by remotely operated stepper motors. The displacement varied pseudo-randomly across trials, allowing to test a large set of configurations. 
To control distal visual cues, experiments were conducted in complete darkness. To support navigation, the track surface was coated with a scentless, washable, nontoxic glow-in-the-dark paint, which provided a faint green glow. A large screen blocked the view of one glowing arm from the other, ensuring the rat could only see the arm it was traversing. Stepper motors operated quietly, and the animals showed no overt behavioral responses to the associated noise or vibrations during track repositioning. 
Each animal was recorded for $7$ days on the track, performing on average about $40$ laps during each run session. The average traversal time ($\sim34$ s) did not significantly differ from pre-training performance on the static track. More details can be found in \cite{eLife}.

\subsection*{Data Analysis}
All analyses excluded epochs when animals ran at speeds $<$2 cm/s. To compare spiking activity in planar and linear reference frames, we used Cartesian 2D coordinates and then linearized the trajectory such that the $x$ coordinate represented distance from one of the food wells, ignoring displacements orthogonal to the track direction. 
Receptive fields were estimated from single-unit spiking activity along the linear track. For each cell, spike positions were binned along the $x$-coordinate and converted into a spatial firing-rate profile by normalizing spike counts with the occupancy, i.e. the time spent by the animal in each spatial position. The resulting firing-rate profiles were smoothed with Gaussian filtering to reduce noise while preserving the structure of place fields. The reception fields were then defined as contiguous regions where the smoothed firing rate exceeded a fixed threshold of 2 Hz above baseline activity. For consistency of analysis, only a single receptive field was retained per neuron: in cases where multiple fields were detected, all but the largest contiguous field containing the peak firing location were discarded. Data analysis was performed in Python.
To quantify the curvature of the receptive field distribution, we followed the exponential model of place-field sizes described by Eq.~(1):
\begin{equation*}
	p(s) = \frac{\xi\, e^{-\xi s}}{e^{-\xi s_{\min}} - e^{-\xi s_{\max}}},
\end{equation*}
where $s$ denotes the field size, $\xi $ is the global curvature parameter, and $s_{\min}, s_{\max}$ are the minimal and maximal observed field sizes. 
To estimate the \emph{local curvature} $\xi_{\mathrm{loc}}$ along the track, we applied this formulation within sliding windows (40 cm) on the spatial axis $x$. In each window, receptive field sizes $s$ with centers located inside the window were extracted, and a maximum likelihood estimate of $\xi$ was computed using the truncated exponential model above. This procedure yielded a spatially resolved profile $\xi_{\mathrm{loc}}(x)$, capturing local variations in field-size scaling. The resulting $\xi_{\mathrm{loc}}(x)$ values were subsequently smoothed with a Gaussian kernel to suppress noise while preserving spatial trends. 
\newpage
\section{Bibliography}
\label{sec:bib}

\end{document}